\documentclass[aps,pra,twocolumn,showpacs,superscriptaddress]{revtex4}
\usepackage{bm,amsmath,amssymb,latexsym,mathrsfs,graphicx,enumerate}
\usepackage[mathcal]{euscript}
\usepackage{hyperref}
\usepackage{epsfig}

\newcommand{\be}{\begin{equation}}
\newcommand{\ee}{\end{equation}}

\begin{document}

\title{\textbf{High frequency polarization switching of a thin
ferroelectric film}}
\author{J.-G.~Caputo$^{1~}$, A.I.~Maimistov$^{2,3}$,E.D. Mishina$^{4}$, E.V. Kazantseva$^{2,4}$,  V.M.~Mukhortov$^{5}$}
\affiliation{\normalsize \noindent
$^1$: Laboratoire de Math\'ematiques, INSA de Rouen, \\
BP 8, Avenue de l'Universite,
Saint-Etienne du Rouvray, 76801 France \\
$^2$: Department of Solid State Physics and Nanostructures, \\
Moscow Engineering Physics Institute,
Kashirskoe sh. 31, Moscow, 115409 Russia \\
$^3$: Department of Physics and Technology of Nanostructures / REC Bionanophysics,\\
Moscow Institute for Physics and Technology, Institutskii lane 9,
Dolgoprudny,
Moscow region, 141700 Russia \\
$^4$: Department of Condensed Matter Physics, Moscow Institute of
Radiotechnics,\\
 Electronics and Automation, Vernadskogo pr. 78, Moscow,
119454 Russia \\
$^5$: Southern Scientific Center, Russian Academy of Sciences,
Rostov-on-Don, 344006 Russia\\
E-mails: caputo@insa-rouen.fr, aimaimistov@gmail.com, elena.kazantseva@gmail.com, mishina57@mail.ru \\
}
\date{\today}

\begin{abstract}
\noindent We consider both experimentally and analytically the
transient oscillatory process that arises when a rapid change in
voltage is applied to a $Ba_xSr_{1-x}TiO_3$ ferroelectric thin
film deposited on an $Mg0$ substrate. High frequency ($\approx 10^{8}
rad/s$) polarization oscillations are observed in the ferroelectric sample.
These can be understood using a simple field-polarization model. In particular
we obtain analytic expressions for the oscillation frequency and the
decay time of the polarization fluctuation in terms of the material
parameters. These estimations agree well with the experimental results.

\end{abstract}

\maketitle


\section{Introduction}

\noindent Unique intrinsic properties make ferroelectric materials
attractive both for fundamental research and applications in devices
using electro-optical, piezoelectric and other effects. Bulk
ferroelectric materials, particularly those based on
barium-strontium titanate $ Ba_xSr_{1-x}TiO_3$ (BST) compounds
\cite{Smolensky,Vendik} are attractive for high-power applications
because of their high dielectric permittivity and small losses.
Ferroelectric-based devices include ultra-fast electrically-controlled phase
shifters for amplitude and phase control. It has been shown that the
dielectric permittivity
of a ferroelectric can be altered by applying an electric field. Therefore
a fast ferroelectric phase shifter controlled by an electric field bias is being
investigated \cite{ferro_switch}
to be used for applications in particle accelerators.
Ferroelectric materials provide significant benefits for several
applications such as switching and control elements. These are able to
handle high peak and average power while maintaining a very short
response time of less than just few nanoseconds. According to some
estimations \cite{Smolensky}, \cite{ferro_switch} the response time
to an applied external electric field is about ~$10^{-11}$ s for
crystalline and ~$10^{-10}$ s for ceramic compounds. The high
permittivity (over 1000) of ferroelectrics makes them potential
candidates to replace silicon oxide dielectrics as storage
capacitors for memory devices. A solid solution barium-strontium
titanate $Ba_xSr_{1-x}TiO_3$ (BST) has a high permittivity and
a composition dependent Curie temperature $T_c$ which varies in a
range of 30 - 400 K. This strong dependence of the dielectric constant on
electric field offers the opportunity to use ferroelectrics as
tunable devices. Ferroelectric based phase shifters organised in an
array have the advantage
of being cheap and consuming a reduced power while
continuously tuning the phase of a high power microwave signal. This
rapid electrical steering is realized by adjusting the bias voltages
on each element. See for example \cite{Vendik_ph_shift_ferroel} and
\cite {Romanofsky} where the phase shifting elements based on BST
thin film capacitors are discussed. There is a certain advantage in
using elements build on thin ferroelectric films because of their
compactness and parameter tunability. The presence of internal stresses
in the thin film ferroelectrics changes their electromechanical and
dielectric properties drastically. Because of this
the dielectric permittivity in a thin film is reduced by about
an order of magnitude compared to one for the bulk media. However
choosing the appropriate substrate
allows to adjust the internal stress level and tune the physical
properties of the thin ferroelectric films.
Thin films of BST deposited by the sputtering technique are discussed in \cite%
{capacitors}, \cite{gatech_capacitors} in reference to the production of compact
tunable capacitors. These elements are attractive for applications in
adaptive impedance matching networks and tunable filters. Size
effects in thin ferroelectric films are discussed in \cite{Fridkin2}.
It is reported that no critical film thickness is required to
obtain polarization switching.

A noteworthy transient effect occurs in
ferroelectric material subjected to an alternating electric field which
causes polarization switching between two stationary states of the
ferroelectric. The switching process is followed by high-frequency
polarization oscillations around their stationary states. Basically
the polarization dynamics of a ferroelectric near its steady state can be
viewed in terms of a damped oscillator with an eigenfrequency determined
by the material parameters. An alternating electric field serves as the
external force that pushes the oscillator away from its equilibrium.
The generation of infrared (IR) radiation by means of polarization
switching was proposed for ferroelectrics \cite{Balkarei}. There the authors
estimate the energy radiated using the dipole approximation.

The study of the transient dynamics of these processes can help
understand how the system parameters can be adjusted to provide
the required transient behavior. That is important for devices
subjected to sharp/shock periodic or aperiodic forces of high
frequency. They then spend most of their time in a transient state,
even if the relaxation time is smaller than the observation period.
Transient processes occurring in ferroelectric might become unwanted
effects if the possible applications require fast switching between
polarization states, such as in FeRAM. In opposite they might be required if
the polarization switching is used to produce oscillations that will
form an IR pulse. The polarization relaxation time and oscillation
frequency are important characteristics of this transient behavior. The
polarization damping constant in ferroelectrics is of the order of
$10^{10} s^{-1}$, as indicated in \cite{Fridkin}. The theoretical
description for an isotropic paraelectric in the framework of the
dynamical Landau-Khalatnikov model being investigated in \cite
{Osman_Ishib_Tilley1,Osman_Ishib_Tilley2} allows to calculate nonlinear
susceptibility coefficients.

Here we discuss an experiment where we observe the
polarization dynamics in a thin BST film.
A similar experiment studying fast polarization switching was
performed earlier by some of the authors. The method of observation
of the polarization uses the
second harmonic generation in the thin film of the BST solid
solution. It is discussed in \cite{Mishina_apl}.
In the present experiment we follow a pump-probe procedure. First
we apply a constant field $E_s$ and
obtain the steady state induced polarization $P_s$. Then we
send in an additional small electric field pulse to probe the film.
In principle this time dependent solution could be described by
the theory set up in \cite{cmk} but we chose in this first study
to describe the relaxation response of the film.  For that we set up
a scattering theory formalism. The bound states of the system $(E,P)$ will
then give the response of the film, ie the typical oscillation
frequency and the radiative decay time. We
identify two channels of damping, the radiative damping and the
inner damping. From the experimental data we estimate the magnitude
of both terms and find that here the inner damping dominates.
There could be other experiments where the contrary happens ie
the radiative damping may dominate.

\section{Model of electromagnetic response in ferroelectric film}

\noindent The theory of ferroelectric has started to develop in the 1930s.
The phase transition theory proposed by Landau
\cite{Landau_JETP}, \cite {Landau_stat1} was applied to describe the
behavior of the ferroelectric near a critical point of the phase
transition. Following the Landau-Ginzburg-Devonshire theory
the thermodynamic Gibbs potential $ \Phi$ in the neighborhood of the
critical point can be represented as a power series of the order
parameter, namely the polarization \cite{Ginzburg45}, \cite{Ginzburg49},
\cite{Ginzburgufn} up to 4-th order, and \cite{Devonshire} up to
terms of order $P^6$.
\begin{equation*}
\Phi= \Phi_0 + \frac{\alpha}{2} P^2 + \frac {\beta}{4} P^4 -E P.
\end{equation*}
The expansion coefficients are $\alpha=a_0(T-T_c)$ and $\beta$. $T_c$
is the Curie temperature. In a solid solution $Ba_xSr_{1-x}TiO_3$ the Curie
temperature depends on the relative concentration $x$ of barium. The
approximate formula to calculate $T_c$ in BST was proposed by \cite{Lemanov}.
It reads $T_c=360x+40$. Particularly $T_c=292 K$ for $x=0.7$. This is the
concentration of barium in the thin film investigated in an experiment on
ferroelectric switching \cite{Mishina_apl}.

We describe an experiment where a thin film of thickness $l$ is
deposited on a dispersionless substrate. The optical properties of
the surrounding media can be characterized by the refractive index
$n(z)$, such that $n(z)=1$ for $z<0$ and $n(z)=n >1$ for $z>0$, i.e.
in the substrate. The electric field $E(z,t)$ is governed by the
Maxwell equation and the polarization obeys the equation of a damped
oscillator driven by the field (compare to \cite{cmk})
\begin{eqnarray}
&& {\partial^2 E \over  \partial z^2} -{n^2 \over c^2}{\partial^2 E
\over  \partial t^2} = \frac {l} {\epsilon_0 c^2}
{\partial^2 P \over \partial t^2}\delta \left( z \right),  \notag \\
[-1.5ex]  \label{sys_eq} \\
&& \frac{\tau^2}{\epsilon_{0}} \frac {\partial^2 P}{\partial t^2}
+\frac{\gamma \tau^2 }{\epsilon_{0}} \frac {\partial P}{\partial
t}+\alpha P + \beta P^3 = E_f(t),  \notag
\end{eqnarray}
where $E_f(t) $ is the electric field inside the thin film. The
constant $\tau$ is proportional to the inverse
of the Born frequency, this value will be estimated for BST later.
Typically the decay constant $1/(\gamma \tau^2) $ is small \cite{Ginzburgufn}.
However we take it into account for generality.

The electromagnetic wave is incident from the left of a film located
at $z=0$. Thus the electric fields outside the thin film
$E^{-}(z,t)$ for $z<0$ and $E^{+}(z,t)$ for $z>0$ are defined by
free wave equations. The time Fourier images of these fields can be written as
\begin{eqnarray*}
&& \tilde{E}^{-}(z,\omega)=A e^{ik_1 z}+B e^{-ik_1z}, \\
&& \tilde{E}^{+}(z,\omega)=C e^{ik_2z } .
\end{eqnarray*}
Here $k_1=\omega/c$ and $k_2=\omega n/c$ are the wave numbers on the left
and on the right of the thin film, $A$ is the Fourier amplitude of
the incident wave, $B$ is the Fourier amplitude of the
reflected wave and
$C$ is the Fourier amplitude of the transmitted wave.

\begin{figure}[tbp]
\centerline{\epsfig{file=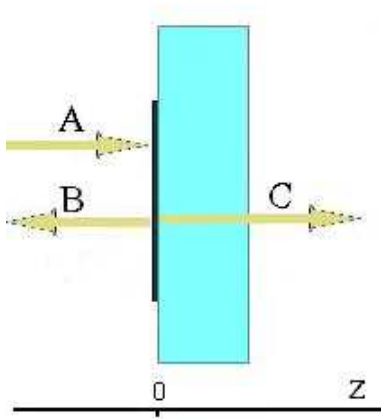,width= 50mm, height=60mm,angle=0}}
\caption{Schematics of the experimental setup where an electric
field pulse polarizes a ferroelectric thin film deposited on a
substrate.} \label{f1}
\end{figure}
According to the boundary conditions at $z=0$ \cite{cmk,Rupasov} the
electric field $\tilde{E}$ and its spatial derivative
$\tilde{E_{,z}}$ are connected by the relations
\begin{eqnarray*}
&&\tilde{E}^{(-)}(z=0,\omega)=E^{(+)}(z=0,\omega),  \\
&&\tilde{E}_{,z}^{(-)}(z=0,\omega)-E_{,z}^{(+)}(z=0,\omega)=
\frac{l\omega^2}{ c^{2}\epsilon _{0}}P(\omega),
\end{eqnarray*}
where $P(\omega)$ is the Fourier image of the thin film polarization.
This leads to the following equations for the Fourier amplitudes
of the left and right electric fields
\begin{equation*}
A+B=C,\hspace{1em}A-B=nC-\frac{il\omega}{c\epsilon _{0}}P(\omega).
\end{equation*}
Hence the amplitude of the transmitted wave $C$ and the amplitude of
the reflected wave $B$ can be expressed through the amplitude $A$
of the incident wave
and the thin film polarization $P(\omega)$:
\begin{eqnarray}
&& C=\frac{2}{n+1}A +\frac{il\omega}{c(n+1)\epsilon _{0}}P(\omega), \label{eq:trans}\\
&&B=\frac{1-n}{1+n}A+\frac{il\omega}{c(n+1)\epsilon _{0}}P(\omega).
\label{eq:reflect}
\end{eqnarray}
Using the inverse Fourier transform we obtain
the amplitude of the electric field inside the thin film $E_f$ as
$E^{+}(z,t)$ at $z=0$ \cite{Rupasov}, i.e.,
$$
E_f(t)=\frac{2}{1+n}E_{in}(t) -\frac{l}{c(n+1)\epsilon _{0}}
\frac{\partial P}{\partial t},
$$
where $E_{in}(t)$ is the electric field of the incident wave. Thus one
can find from the second equation of the system (\ref{sys_eq}) that
the polarization of the ferroelectric thin film is governed by the
following equation
\begin{eqnarray}
&&\frac{\tau^2}{\epsilon_0}\frac{\partial^2 P}{\partial^2
t}+\left(\frac{\gamma\tau^2 }{\epsilon_{0}} + \frac{l}{c(n+1)\epsilon
_{0}}\right)\frac{\partial P}{\partial t} +
\notag \\
&& \qquad ~~~~ + \alpha P + \beta P^3 =\frac{2}{1+n}E_{in}(t).
\label{eq:polarizat}
\end{eqnarray}

This expression shows that two relaxation channels exist. One of
them is the ordinary one related with inner friction i.e., the
$\gamma$-term.  The second relaxation channel is due to the
radiation process. The variation of the polarization generates the
electromagnetic field outside the film.

The polarization of the thin film evolves according to
(\ref{eq:polarizat}) as an oscillator with damping  and a forcing that is
equal to $E_s = 2E_{in}(t)/(1+n)$. Furthermore, this equation allows
to consider $E_{in}$ as a constant electric field i.e. a constant
voltage. In this case $E_s =V_s/l$ where $V_s$
is the applied voltage.

\subsection{Relaxation to a steady state polarization}

In the absence of an external electric field the ferroelectric
possesses two equilibrium states, each corresponding to different
polarities. In the paraelectric phase there is only one equilibrium
corresponding to an unpolarized state. The experiment under
consideration was performed at room temperature which is above the
Curie temperature for BST so that the sample is in the paraelectric
state. However there is another way to change the polarization
state. This can be induced by applying an external field or a
stress. In this experiment we chose to do the former. The induced
steady state polarization $P_s$ is defined as the fixed point of the
equation (\ref{sys_eq}) where we assume that the electric field
$E_f(t)=E_s$ is constant and the polarization $P=P_s$ are time
independent. We get
\begin{equation}
\alpha P_{s}+\beta P_{s}^{3}=E_s . \label{steady_state}
\end{equation}

We assume the solution of (\ref{sys_eq}) to be of the form
\begin{equation}\label{ep_pert}
E = E_s + \delta E, ~~~P = P_s +\delta P,
\end{equation}
where $\delta E$ and $\delta P$ are small compared respectively to
$E_s$ $P_s$. Plugging this into the equations (\ref{sys_eq}) we
get the linearized equations around the stationary solution $(E_s,P_s)$
\begin{eqnarray}
&&
{n^2 \over c^2}{\partial^2 \delta E \over \partial t^2}
-{\partial^2 \delta E \over \partial z^2}
= -\frac {l} {\epsilon_0 c^2 } \frac {\partial^2 \delta P}{
\partial t^2}\delta \left( z \right),  \notag \\
[-1.5ex]  \label{ep_lin} \\
&& \frac{\tau^2}{\epsilon_{0}} \frac {\partial^2 \delta P}{\partial
t^2} +\frac{\gamma\tau^2 }{\epsilon_{0}} \frac {\partial \delta
P}{\partial t} +(\alpha  + 3\beta P_s^2)\delta P  = \delta E(t).
\notag
\end{eqnarray}

\subsection{Scattering of linear waves by the thin film}

We now proceed to solve the linearized equations (\ref{ep_lin})
by using a scattering theory formalism.
We separate time and space by assuming a periodic solution
\be\label{ep_periodic}
\delta E = e(z) e^{-i\omega t},~~~\delta P = p e^{-i\omega t}.
\ee
We get
\begin{eqnarray}
&& n^2 k^2 e + {\partial^2 e \over \partial z^2}
= -\frac {l}{\epsilon_0} k^2 \delta \left( z \right) p ,  \notag \\
[-1.5ex]  \label{ep_z} \\
&& (\Omega^2 + i\gamma\tau^2 \omega - \tau^2 \omega^2 ) p = \epsilon_{0} e(0),  \notag
\end{eqnarray}
where the wave number $k = \omega / c$ and where we introduced
\be\label{big_omega} \Omega^2 = \epsilon_{0}  (\alpha  + 3\beta
P_s^2).\ee

In the scattering we assume the electromagnetic wave to be incident
from the left of the film located at $z=0$. We then have
\be\label{scat} e =  e^{ikz} + R e^{-ikz} ~, z<0~;~~~~ e = Te^{iknz
}, z>0~~,\ee where $R$ is the amplitude of the reflected wave and
$T$ the amplitude of the transmitted wave. These expressions are
related to the $A,~ B, ~C$ parameters of the previous section through
the relations $R= B/A,~T=C/A$. We have the following
interface conditions at $z=0$ \cite{cmk,Rupasov}
\be\label{interface} e(0^-)= e(0^+),~~-[e_z]_{0^-}^{0^+} =k^2 {l
\over \epsilon_0}p.\ee They imply
$$ 1+R = T,~~ n T -(1-R)= -i k {l \over \epsilon_0}p$$
Using the second relation of (\ref{ep_z}) to obtain $p$ we get the
transmission coefficient $T$ \be\label{trans} T = {2c (\Omega^2 +
i\gamma\tau^2 \omega - \tau^2 \omega^2) \over c(n+1)(\Omega^2
+ i\gamma\tau^2 \omega - \tau^2 \omega^2) + i  l \omega}.\ee The refraction
coefficient is \be\label{refl} R = {c (1-n) (\Omega^2 + i\gamma\tau^2
\omega - \tau^2 \omega^2) - i l \omega\over c(1+n)(\Omega^2 + i\gamma\tau^2
\omega - \tau^2 \omega^2) + i  l \omega}.\ee The bound states
are the poles of the reflexion and transmission coefficients. Their
existence indicates that the system has resonant modes that can be
excited by an incoming wave. The real part of the bound states is
the oscillation frequency and the imaginary part is the inverse of
the decay time of the mode.

The poles of $R,T$ are given by
$$c(n+1)(\Omega^2 + i\gamma\tau^2 \omega - \tau^2 \omega^2) + i l\omega =0,$$
which is the second degree equation \be\label{eq_wp}
\omega^2 \tau^2
-i\omega \left({l \over (n+1) c}+\gamma\tau^2 \right)  -\Omega^2
=0,\ee whose roots are
\be\label{wp}
\omega = {i \over 2  }
\left ({l \over  (n+1) c} +\gamma\right) \mp
\sqrt{{\Omega^2 \over \tau^2} -
{1\over 4 } \left [{l \over  (n+1) c} +\gamma \right ]^2 }.\ee The
imaginary part and real part of $\omega = \omega_r + i\omega_i$ give
respectively the decay time $T_{dec}$ and the oscillation period
$T_{osc}$ \be\label{decay_osc} T_{dec} = {1 \over
\omega_i},~~~T_{osc} = {2 \pi \over \omega_r} .\ee We will estimate
these parameters for the experiment in the next section.

\subsection{Green function solution of the linearized equation}

The linearized equation
\be\label{ptt} \frac{\tau^2}{\epsilon_{0}}
\frac {\partial^2 \delta P}{\partial t^2} +\frac{\tau^2 \gamma }{\epsilon_{0}} \frac {\partial \delta P}{\partial t} +(\alpha
+ 3\beta P_s^2)\delta P  = \delta E(t) , \notag \ee can be solved to
obtain the polarization response $\delta P(t) $ to a give incoming
electric field $\delta E(t) $. For that we introduce the Green
function $G(t-t_0)$ which satisfies \be\label{green_t} \tau^2 \frac
{\partial^2 G}{\partial t^2} +\tau^2 \gamma \frac {\partial G}{\partial
t} +\Omega^2 G  = \epsilon_0 \delta (t-t_0) . \ee Using the Laplace
transform
$${\hat G}(s) \equiv \int_0^\infty e^{-s t} G(t) dt,$$
and assuming that $G(0)=0$ and $\partial G/\partial t(0)=0$ we
obtain \be \label{g_hat} {\hat G}(s) = { \epsilon_0 e^{-s t_0}\over
\tau^2 s^2 + \tau^2 \gamma s + \Omega^2}. \ee To get the inverse Laplace
transform one expands this rational function as
$${\hat G}(s) = {\epsilon_0 \over 2(s_1-s_2)} \left (
{e^{-s t_0} \over s-s_1 }
- {e^{-s t_0} \over s-s_2 } \right ),$$
where $s_1$, $s_2$
are the roots of the denominator of (\ref{g_hat}).
This yields the Green function
\begin{eqnarray}
&&G(t-t_0)  = { \epsilon_0 \over 2(s_1-s_2)} \left[  e^{s_1(t-
t_0)}- e^{s_1(t- t_0)} \right],
~~  t>t_0 \notag \\
 &&G(t-t_0)  = 0,
 ~~~~~~~~~~~~  t< t_0 \notag
\end{eqnarray}

The roots are complex conjugate $s_1 = \omega_i + i \omega_r$
,~$s_2=\omega_i - i \omega_r$ so we obtain the final result
\begin{eqnarray}
&&G(t-t_0) = e^{\omega_i t} {\sin[\omega_r(t-t_0)] \over 2 \omega_r},
 ~~ t>t_0  \notag \\
&& \label{green}\\
&&G(t-t_0)= 0 \qquad \qquad \qquad \qquad ~
t < t_0 \notag
\end{eqnarray}
The polarization response $\delta P(t) $ to a given perturbation of
the electric field $\delta E(t)$ is the convolution integral
\be\label{pol_resp} \delta P(t) = \int_{-\infty }^t G(t-t_0) \delta
E(t_0) dt_0.\ee

\subsection{Induced polarization caused by a short electric pulse}

Now let us consider an alternative method to investigate the
polarization response. Suppose that a ferroelectric film is in a
polarized state $P_s$ caused by a constant electric field $E_s$.
At certain moment we send in a short electric pulse so
the polarization of the film changes during a
short time period.
After that the polarization relaxes to the steady state
position $P_s$. If the electric pulse is sufficiently short i.e.
close to a "$\delta$ function" then the polarization response
will follow the Green function (\ref{green}). We term this action
"$\delta$-function-like pushing"\, the nonequilibrium polarization.

The linearization of the equation for the polarization (\ref{eq:polarizat})
near the steady state $P_{s}$ assuming
$P=P_{s}+p$ with the condition $p<<P_{s}$ results in the
equation for the Green function (\ref{green_t}). Let us suppose that the
extremely short electric pulse acts at $t=0$. Then we can conclude
that the evolution of $p(t)$ after $t=0$ is described by
$$
p(t) \sim e^{\omega_i t} {\sin[\omega_r t)] \over 2 \omega_r}.
$$
Hence the polarization decay rate is defined by the imaginary part of the
complex frequency (\ref{wp}), $\omega_i$ and the corresponding time
for the polarization to attain the equilibrium is
\begin{equation*}
T_{dec}=1/\omega_i.
\end{equation*}
The real part of the frequency (\ref{wp}) corresponds to oscillations of
the polarization as it is approaches the steady state value. Here we recover the
results obtained using the scattering formalism.

\section{Experimental results}

The experiment was performed using the nonlinear optical
stroboscopic technique as in
\cite{Mishina_apl}. For the second harmonic (SH) generation, the radiation
of a titanium-sapphire laser (MaiTai, New-Port-SpectraPhysics) was
used with a pulse duration of 100 fs, a wavelength of 780 nm, a repetition
rate of 100 MHz and an average power of 100 W. The experiment was
performed at room temperature. The $70 nm$ thick
$Ba_{0.7}Sr_{0.3}TiO_3$ films were deposited onto a $MgO$ substrate
by RF sputtering. For such a composition, the Curie temperature equals
$Tc=20$ C. However for thin films the phase transition is blurred
around this value and this is confirmed by the presence of a narrow
hysteresis above $T_c$ \cite{Mishina_apl}. A voltage pulse of duration about
$25$ ns, produced by an Avtech pulse generator, was applied to the
copper contacts on the BST film. The polarization response of the
ferroelectric film was measured as the coherent
SH intensity \cite{experiment}.

\begin{figure}[tbp]
\centerline{
\epsfig{file=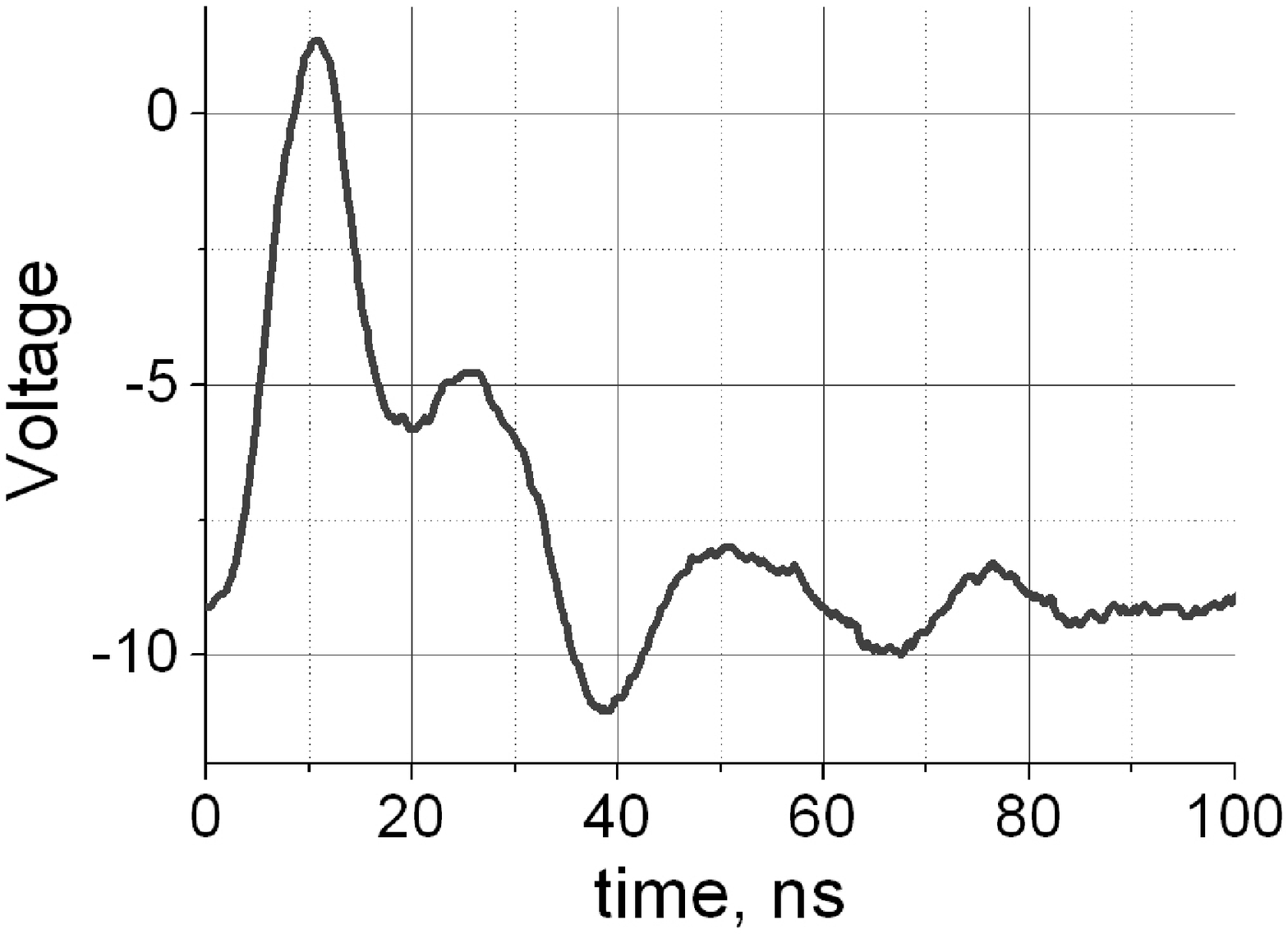,width= 40 mm, height=49.2 mm,angle=0}
\epsfig{file=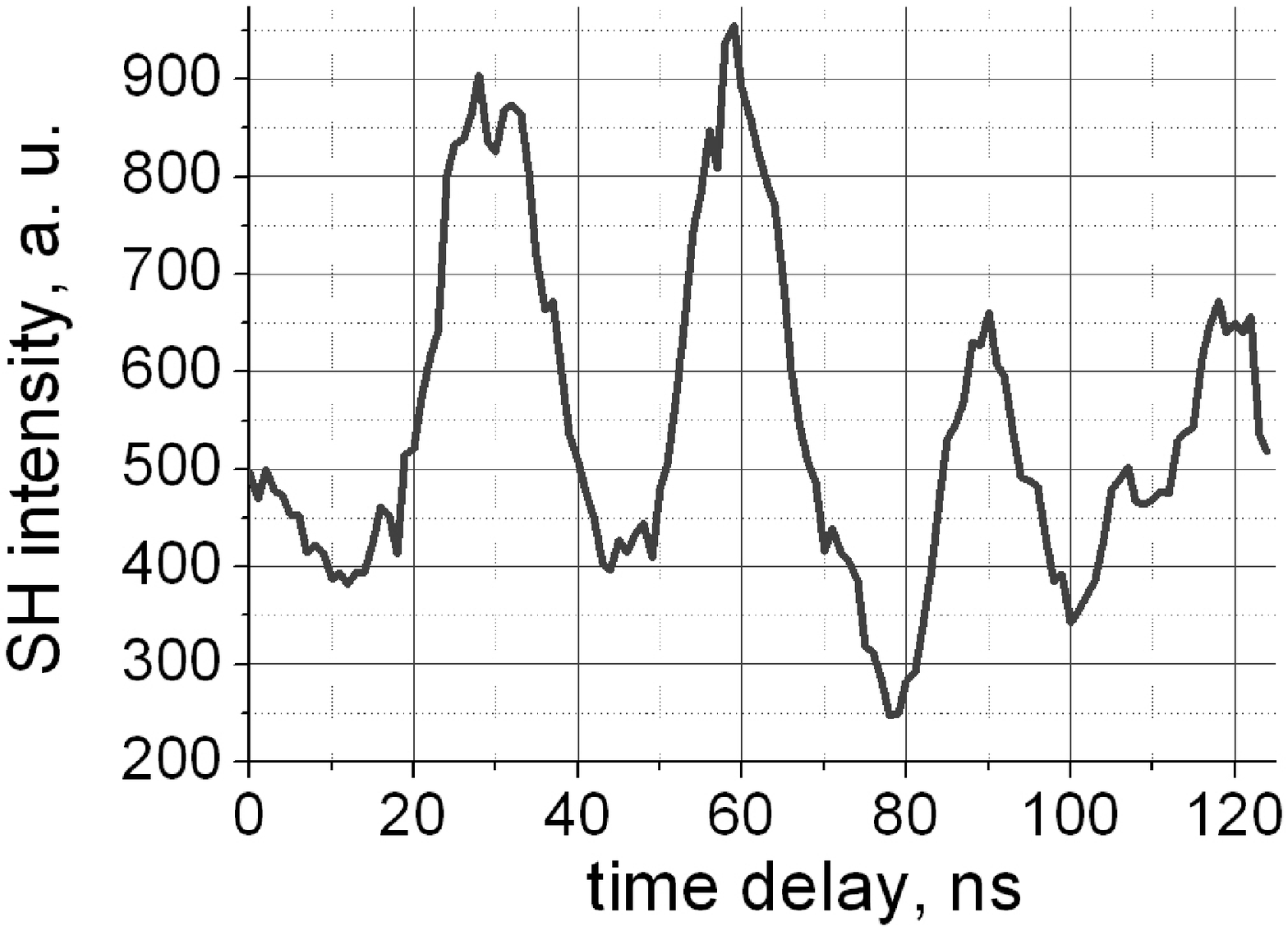,width= 40 mm, height=49.2 mm,angle=0}}
\caption{Plots of a $'zero'$ electric pulse as a function of time (left panel) and the subsequent polarization oscillations of the film as a function of time (right panel). The constant field
$E_s$ is negative.
} \label{fig:exp1}
\end{figure}

\begin{figure}[tbp]
\centerline{
\epsfig{file=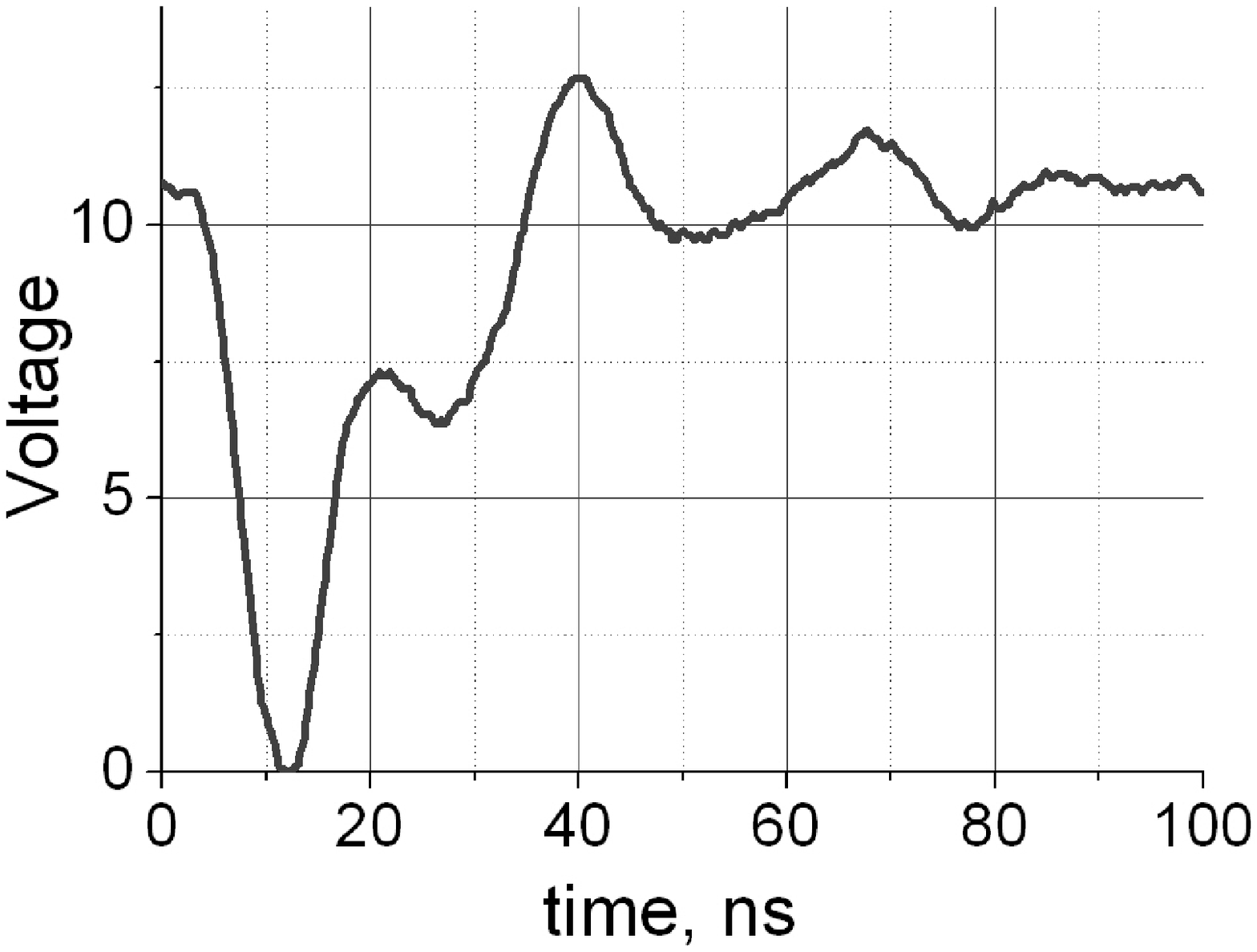,width= 40 mm, height=49.2 mm,angle=0}
\epsfig{file=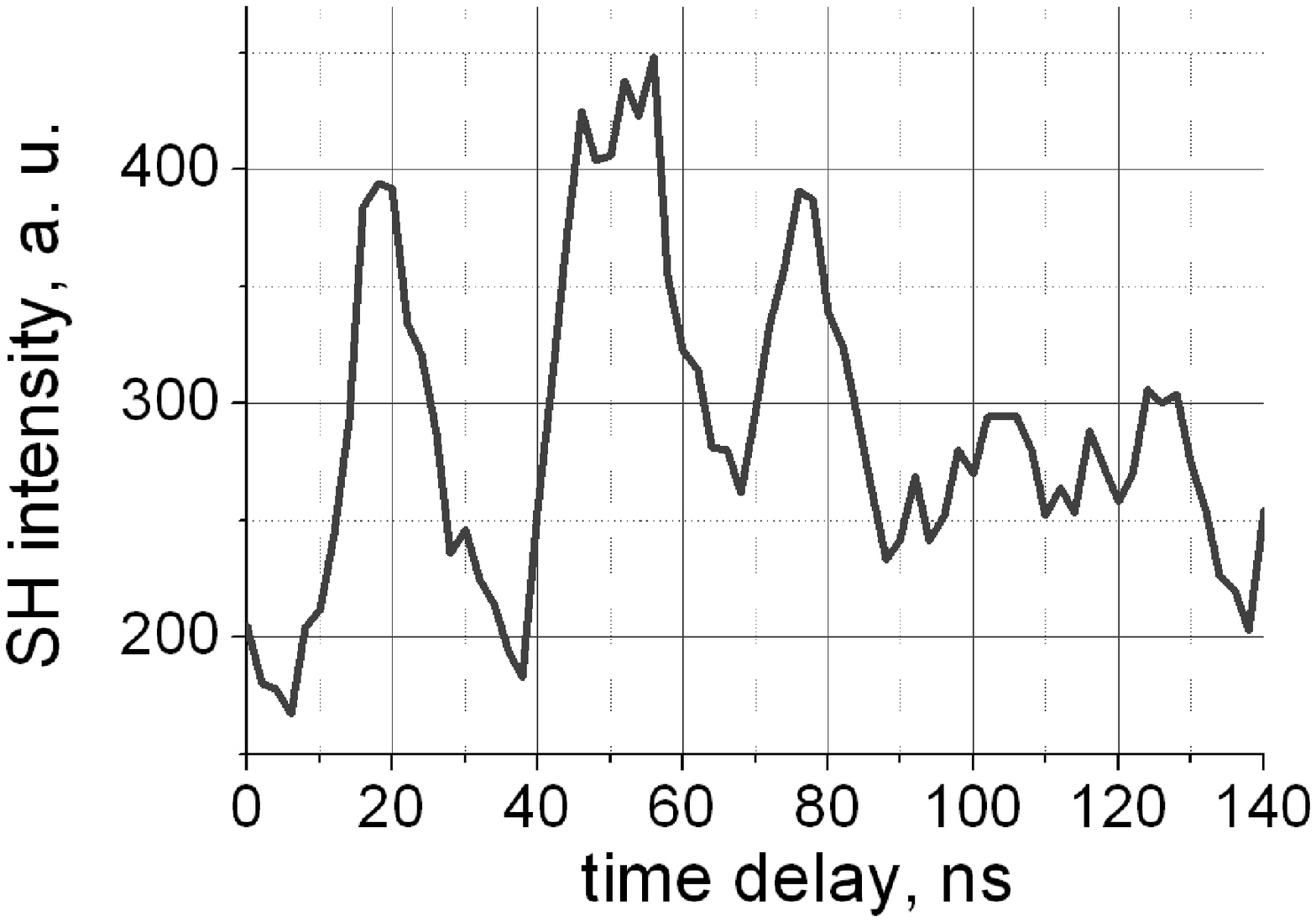,width= 40 mm, height=49.2 mm,angle=0}}
\caption{Plots of a $'zero'$ electric pulse and the subsequent polarization
oscillations of the film as in Figure \ref{fig:exp1} except that the
constant field $E_s$ is positive.
} \label{fig:exp2}
\end{figure}

Figures \ref{fig:exp1} and \ref{fig:exp2} show the
polarization response as a function of time
of the BST thin film to electric pulses
having the same amplitudes and opposite polarities. The left panels
show the incoming electric field pulses as a function of time and
the right panels show the SH intensity (proportional to the
square of the polarization) as a function of time. In figure
\ref{fig:exp1} the electric field pulse (left panel in the figure) is
realized by a rapid spiking to a zero value from a constant
negative voltage background with return to the same constant
negative value. For figure \ref{fig:exp2} the electric pulse
(left panel at the figure) drops from its constant positive value to
zero and returns to the constant. In both measurements the spiking
pulse will be called the 'zero' pulse. The pulse profile is
intended to have a narrow bell shape, however because of certain
setup drawbacks a low-amplitude tail appears.
In the right panels of Figures \ref{fig:exp1} and \ref{fig:exp2}
it can be seen that the polarization oscillates long after
the 'zero' electric pulse has passed through the thin film. Then the
eigenfrequency of the polarized thin film can be determined.

In both pictures the polarization oscillates around its stationary
values (defined by the constant background electric field) with very
close oscillations periods, about
$60$ ns. This agrees with the relation (\ref{osc_period}) in the limit
when the second term can be neglected in the square root because
$\Omega$ only depends on the amplitude $|E_s|$.

The next figure (\ref{fig:exp3}) shows the
polarization dynamics when there is no constant electric field
$E_s$ after a delta function like electric pulse of
negative polarity (shown in the left panel)
passes through the film. This type of electric pulse is the analog
of a 'normal pulse' studied experimentally in
\protect\cite{Mishina_apl}.
When the film is not polarized the SH cannot be generated, so that its intensity
is about zero. The SH signal from the perturbed unpolarized
state is a few times smaller than the one for the previous experiments.
There the steady polarized state of the thin film was studied by a
'zero' pulse. There are almost no polarization oscillations when
perturbing a non polarized thin film. Again this agrees with the
estimate (\ref{osc_period}).

\begin{figure}[tbp]
\centerline{
\epsfig{file=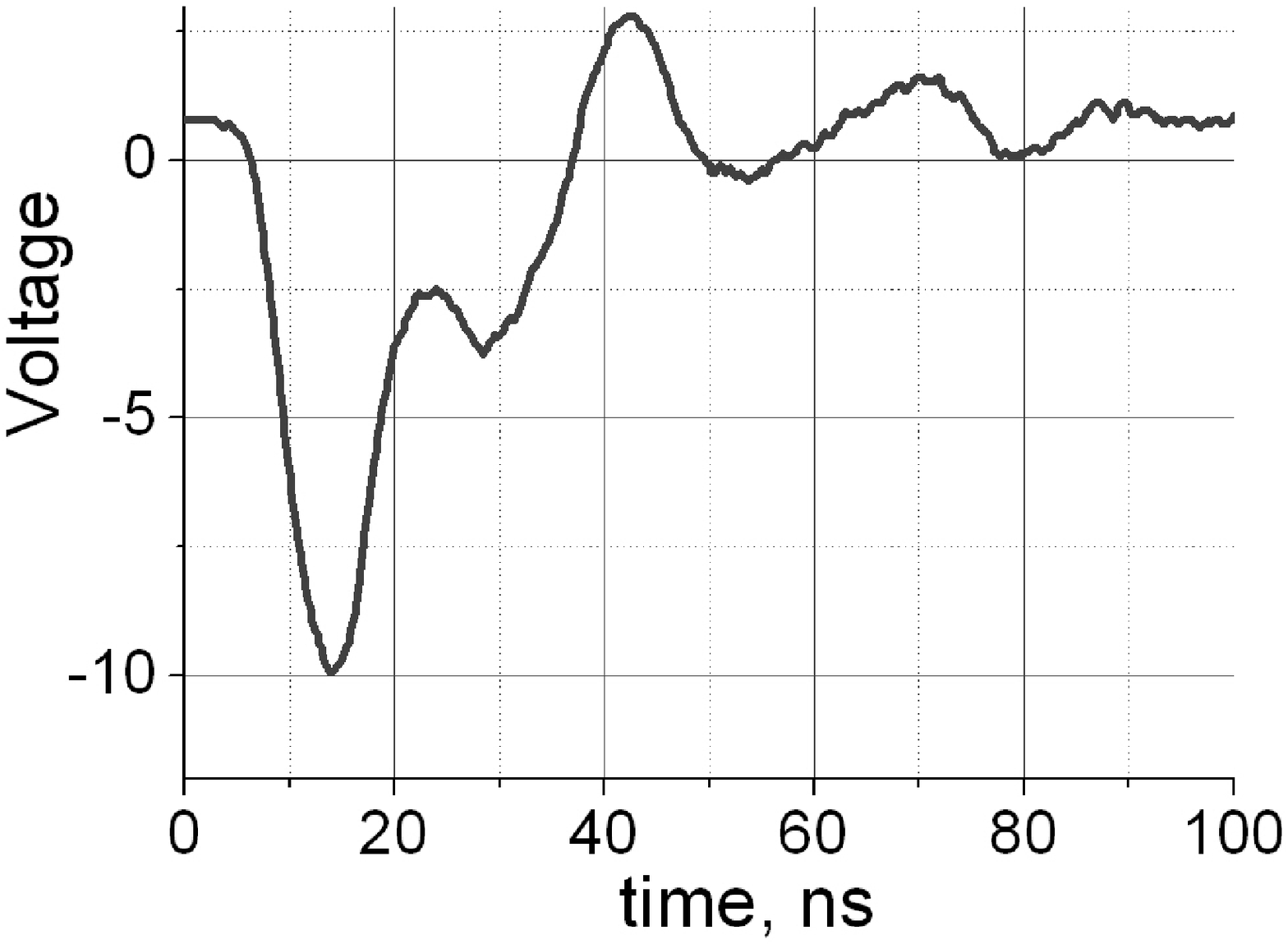,width= 40 mm, height=49.2 mm,angle=0}
\epsfig{file=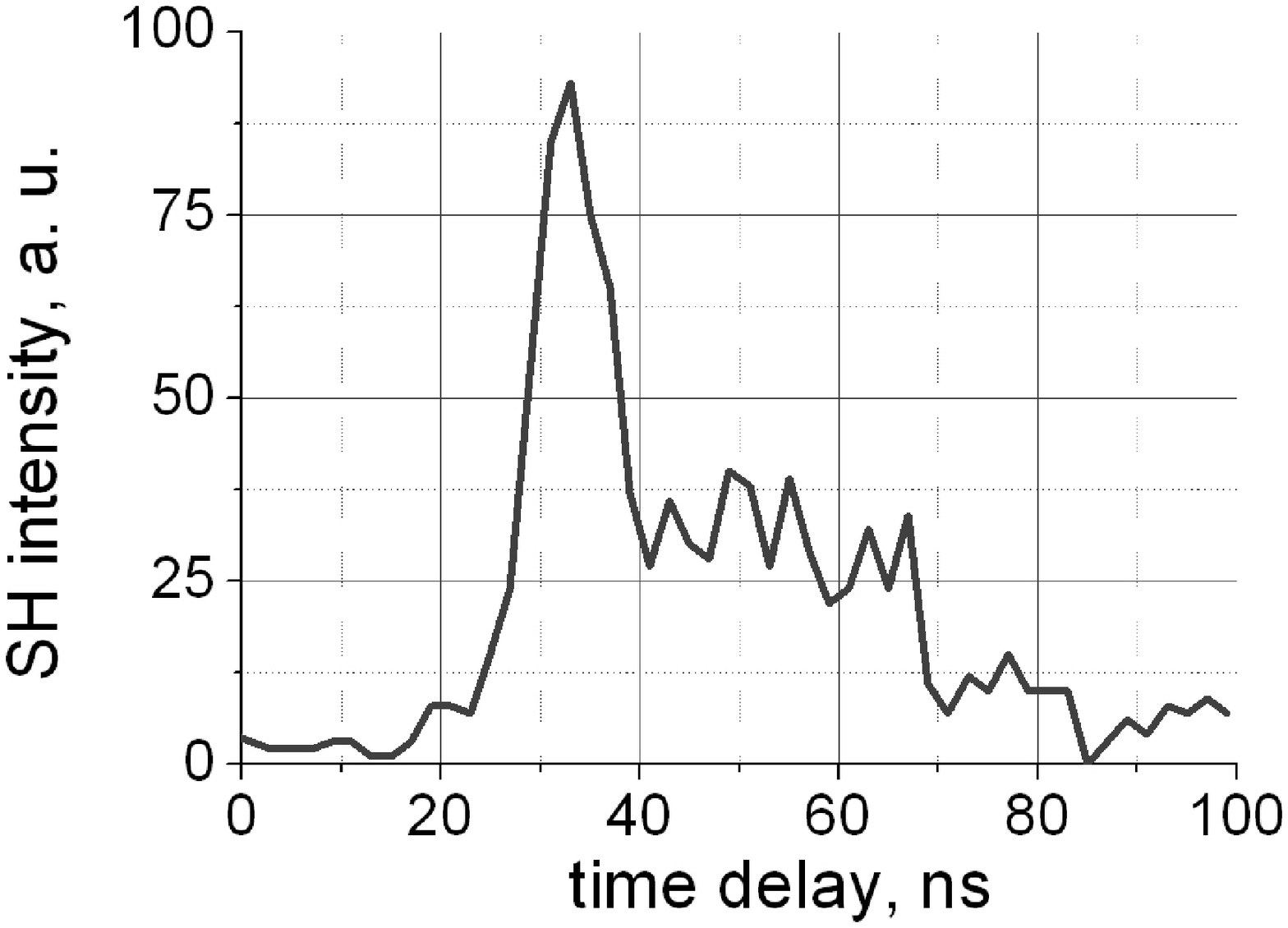,width= 40 mm, height=49.2 mm,angle=0}}
\caption{Left panel, plot of the electric field pulse as a function
of time for a "normal pulse" (see text). The right panel shows the
subsequent polarization of the thin
film as a function of time.
}
\label{fig:exp3}
\end{figure}

\subsection{Discussion of the experimental results}

There is a relatively broad range of experimentally
obtained values of the Landau coefficients in BST solid solutions.
The numbers vary with the fabrication method.  The Landau coefficients
found by \cite{Alpay_jap06} for a BST solid solution $Ba_{0.7}Sr_{0.3}TiO_3$
in SI units are presented in Table I.
\begin{table}
\begin{tabular}
{| c | c | c |}
  \hline
  \hline
  $ T_c, ~[C]$ &  $ \alpha, [m/F] $ & $ \beta, [ m^5/C^2F] $ \\
     \hline
  34 & $2.2\cdot 10^6$ & $2.52\cdot 10^8$ \\
  \hline
  \hline
 \end{tabular}
\caption{Material parameters for $Ba_{0.7}Sr_{0.3}TiO_3$ .}
\end{table}
In the experiment performed the film thickness is $l=70 nm$
(as in \cite{Mishina_apl}).
We assume that the substrate refraction index is $n=1.5$.
These parameters enable to calculate the frequency $\Omega$.
For a normal pulse the applied voltage $V_s=0$ so
$E_s=0$ and $P_s=0$. This gives
$\Omega^2 = \epsilon_0 \alpha \approx 2 10^{-5}$.
In the presence of an electric field $E_s= V_s /l $ where $V_s = 10
Volts$ we have $E_s \approx 10^8 V/ m$ so that from (\ref{steady_state})
we get $P_s \approx 0.79 C/m^2$. This gives $\Omega^2 = \epsilon_0
\alpha \approx 1.8 10^{-3}$.

First let us estimate the term $ l [(n+1) c]^{-1}$ in the formula
(\ref{wp}). We have $ l [(n+1) c]^{-1} \approx 10^{-16}$ which is
very small so that this radiative damping can be completely
neglected for this particular experimental situation.
\begin{table}
\resizebox{0.5\textwidth}{!}{%
\begin{tabular}
 {|c|c| c|c |c| } \hline \hline
  & $\gamma=10^6 [s^{-1}]$  & $10^7$ & $10^8$ & $10^9$  \\ \hline
$\tau= 10^{-6} [s]$   &   $10^{-6}$& $10^{-7}$& $10^{-8}$& $10^{-9}$\\ \hline
 $10^{-7}$   & $10^{-6}$& $10^{-7}$& $10^{-8}$& $10^{-9}$ \\ \hline
$10^{-8}$& $1.4 \cdot 10^{-6}$& $10^{-7}$& $10^{-8}$& $10^{-9}$ \\ \hline
 $10^{-9}$& $2\cdot 10^{-6}$& $1.4 \cdot 10^{-7}$& $10^{-8}$& $10^{-9}$ \\ \hline
 $10^{-10}$& $2\cdot 10^{-6}$& $2\cdot 10^{-7}$& $1.4\cdot 10^{-8}$& $10^{-9}$ \\ \hline
 $10^{-11}$& $2\cdot 10^{-6}$& $2\cdot 10^{-7}$& $2\cdot 10^{-8}$ & $1.4\cdot 10^{-9}$ \\ \hline
 $10^{-12}$& $2\cdot 10^{-6}$& $2\cdot 10^{-7}$& $2\cdot 10^{-7}$ & $2\cdot 10^{-9}$ \\ \hline
 \end{tabular}}
\caption{The decay time $T_{dec}$ as a function of the parameters
$(\tau,\gamma)$ for the zero pulse $V_s=0$ volts.}
\label{tab_decay}
\end{table}
>From the experimental data it can be seen that
the polarisation response for a voltage $V_s = 0$ is qualitatively
different
from the one for $V_s = 10 $ volts. In particular it has a smaller
decay time and practically no oscillations. Let us estimate the
decay time $T_{dec}$ from (\ref{wp}). We have
$$T_{dec} = 2/ \gamma $$
if $(\Omega^2 /\tau^2 - \gamma^2/4 ) >0$, otherwise
$$T_{dec} = \left({\gamma \over 2} +
\sqrt{{\gamma^2 \over 4}-{\Omega^2 \over \tau^2}}\right)^{-1} . $$ This
decay time is shown in Table II for different values of the free
parameters $(\tau,\gamma)$. Clearly if $(\Omega^2 /\tau^2 - \gamma^2/4
)
 >0$ there
will be oscillations in the polarization. Comparing the estimates of
the table with the result shown in Fig. \ref{fig:exp3} indicates
that $\gamma \approx 10^7 s^{-1}$. This agrees with the value
obtained in a previous study by the authors \cite{experiment}.

\begin{table} \label{tab:osc}
\resizebox{0.5\textwidth}{!}{%
\begin{tabular}
 {|c|c| c|c |c| }
  \hline
  \hline
           & $\gamma=10^6~[s^{-1}]$  & $10^7$ & $10^8$ & $10^9$  \\
  \hline
$ \tau=10^{-6} ~[s]$ &  0&  0&  0&  0 \\ \hline
$10^{-7}$& $1.5 \cdot 10^{-5}$& 0& 0&  0 \\ \hline
$10^{-8}$&  $9.6\cdot 10^{-7}$&  $1.5\cdot 10^{-6}$&  0&  0\\ \hline
$10^{-9}$&  $9.6\cdot 10^{-8}$&  $9.6\cdot 10^{-8}$&  $1.5\cdot 10^{-7}$&  0\\ \hline
 $10^{-10}$&  $9.6\cdot 10^{-9}$&  $9.6\cdot 10^{-9}$&  $9.6\cdot 10^{-9}$&  $1.5\cdot 10^{-8}$\\ \hline
 $10^{-11}$&  $9.6\cdot 10^{-10}$&  $9.6\cdot 10^{-10}$&  $9.6\cdot 10^{-10}$&  $9.6\cdot 10^{-10}$\\ \hline
$10^{-12}$&  $9.6\cdot 10^{-11}$&  $9.6\cdot 10^{-11}$&  $9.6\cdot 10^{-11}$&
$9.6\cdot 10^{-11}$\\
\hline
 \end{tabular}}
\caption{The oscillation period $T_{osc}$ as a function of the parameters
$(\tau,\gamma)$ for the non zero pulse $V_s=10$ volts.}
\end{table}

Let us now compare the oscillation period obtained for the non zero
pulse $V_s = 10$ volts with the estimate (\ref{wp}). We have \be
\label{osc_period} T_{osc} = 2 \pi \left( \sqrt{ {\Omega^2 \over
\tau^2} -{\gamma^2 \over 4 }}\right)^{-1}. \ee This value is
computed for different values of $\gamma$ and $\tau$ and reported in
Table III. The zero entries correspond to the situation where
$\Omega^2 / \tau^2 -\gamma^2/4 <0$. As can be seen the values closest to what is
observed in the experimental plots Figs. \ref{fig:exp1} and
\ref{fig:exp2} are $\tau=10^{-10} s$ and $\gamma=10^7$. This value
of the oscillation  time $\tau$ agrees with the one that can be
computed using the speed of sound $c_s = 5 10^3 m/s$ in the material
and the film thickness $l$. We get $\tau = l/c_s\approx 10^{-11} s$.
The estimate for the bulk material using the Newton equation
describing the lattice oscillations $\tau^2_{hf} \approx M \epsilon_0
/(N e^2) \approx 10^{-26} s^{-2}$ where $M$ is the atomic mass of
BST gives $\tau \approx 10^{-13} s$ which is much too small.

\begin{figure}[tbp]
\centerline{
\epsfig{file=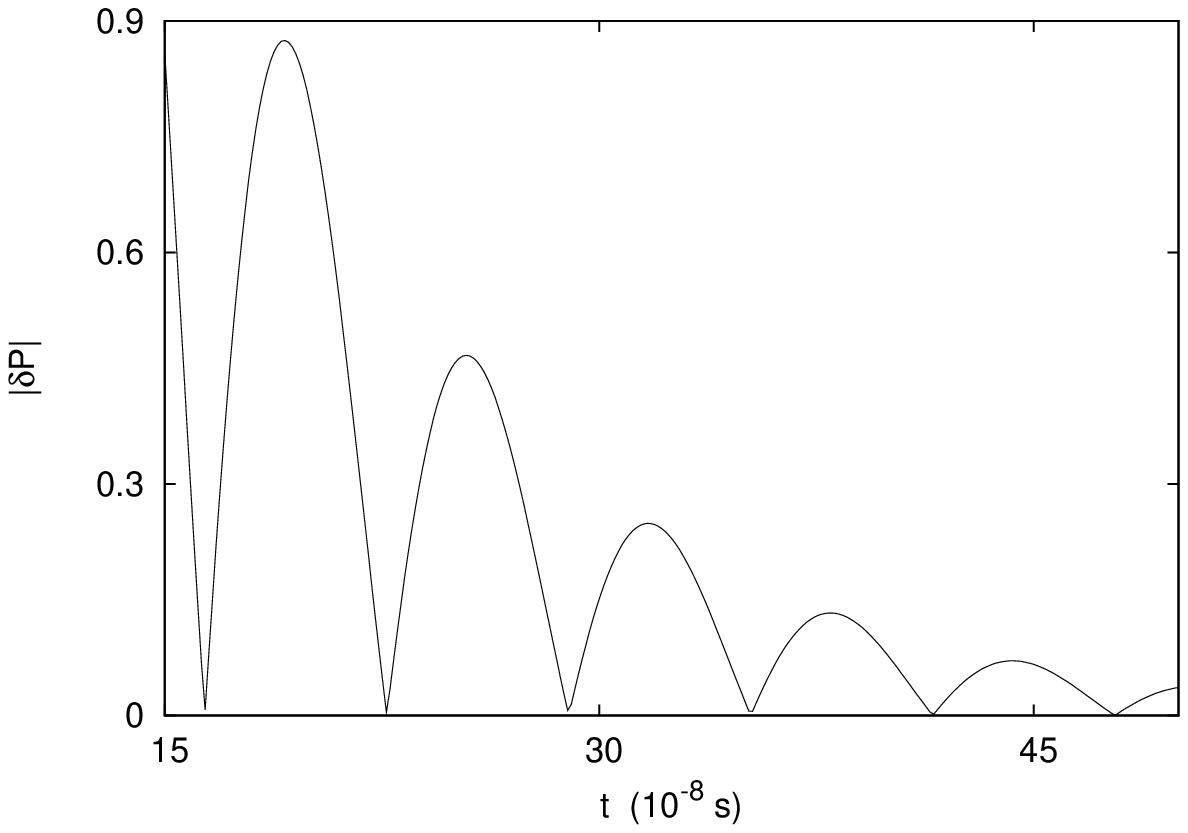,width= 45 mm, height=60mm,angle=0}
\epsfig{file=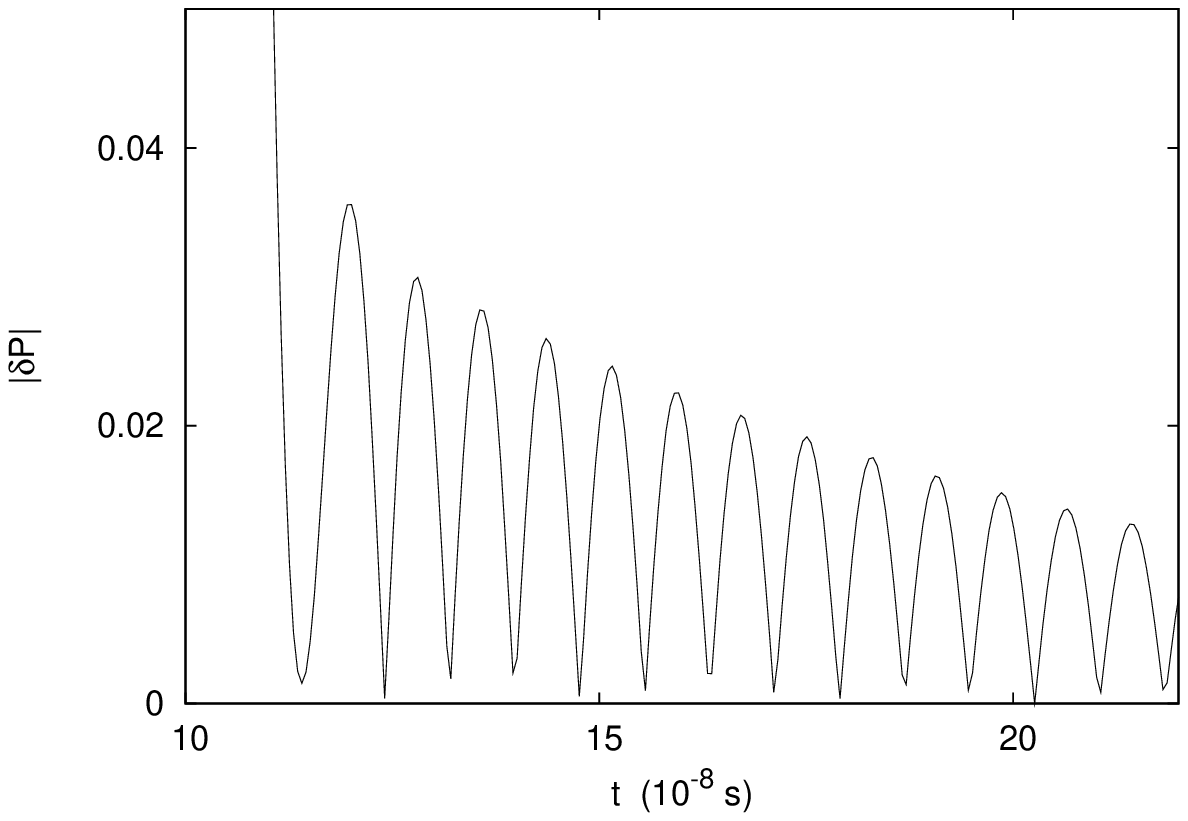,width= 45 mm, height=60mm,angle=0}}
\caption{The polarization response of the film for a normal pulse
($E_s=0$) (left panel) and for a zero pulse (right panel). }
\label{fig:pola_resp}
\end{figure}

We have used the expression (\ref{pol_resp}) to compute the polarization
response of the system to a given perturbation $\delta E$ of the
electric field. We assume a gaussian $\delta E$
\be\label{delta_e}
\delta E(t) = \exp \left (-{t^2 \over 2w_e}  \right ) , \ee
and normalize times by $T = 10 ns$. We chose  $\gamma=10^7 s^{-1},~
\tau=10^{-10} s$. In the normalized units we have $\gamma= 0.1,~
w_e= 1 $. The normalized frequency $\omega = 0.45$ for the zero pulse
and $\omega = 4.5$ for the normal pulse. We calculated the
polarization response by numerical integration using the trapeze
method. To compare with the experimental data we computed
$\delta P^2$.
The results are presented in Fig. \ref{fig:pola_resp}
for a normal pulse ($E_s=0$) in the left panel and for
a zero pulse in the right panel. One can see on the left
side of the plots the first response of
the polarization. We have omitted it because it is the
forced response due to the probing pulse $\delta E$.
We only present the subsequent free evolution of $\delta P$.
For $V_s=0$ the polarisation decays with few oscillations.
For $V_s=10 volts$ the polarization oscillates much longer.
This is in quantitative agreement with the experimental plots
presented above.

\section{Conclusion}

We analyzed the
polarization oscillations occurring in a thin ferroelectric
film as a short electric pulse crosses it. We consider that
the ferroelectric is in the paraelectric phase (high temperature).
As in a pump probe experiment the film is initially in a static
polarization state $P_s$ induced by a constant voltage $V_s$.
Using the Landau-Ginzburg-Devonshire theory
we computed this static polarization and the subsequent oscillations
of the polarization induced by a short voltage pulse. These were
analyzed using a scattering theory formalism and a Green's function
approach. Two channels of dissipation were identified, a radiative damping
and an intrinsic damping.

This theory was applied to explain the experimental time evolution of the
polarization for a thin
ferroelectric film of BST. The polarisation is estimated indirectly through
second harmonic generation. For this experimental situation we show that
the radiative damping can
be neglected and only the intrinsic damping should be considered.
From a comparison of our model to the experimental
plots we estimated the important parameters $\tau$ the response time
and $\gamma$ the relaxation coefficient. Using these values our theoretical
estimates of the decay time and of the oscillation period agree well
with the observations. In particular for a normal pulse for which $V_s=0$
the damping dominates and we only see a few oscillations of the
polarization. On the contrary for a zero pulse for which $V_s=10 volts$
the damping time is longer compared with the oscillation period.
Here we obtain many oscillations of the polarization.

Although the radiative damping appears as artificial in this particular
high temperature situation, at low temperature phonons become frozen
and the radiative damping may become predominant.
The model can be further elaborated by including the spatial inhomogeneity of
the thin film, the depolarization effects of the boundaries, and taking
into account the
internal stresses of the film. Also the dynamics of the polarization of
a thin film multilayered structure can be investigated by generalizing
the model.

\section{Acknowledgments}

JGC thanks the Centre de Ressources Informatiques de Haute-Normandie
for the use of its computing facilities. AIM is grateful to the
Laboratoire de Math\'ematiques, INSA de Rouen for hospitality and
support. This research was supported by RFBR grants No.
09-02-00701-a and 09-07-12144-ophi.

\end{document}